\documentclass[usenatbib]{mn2e}
\usepackage{graphicx, amssymb, aas_macros, natbib}
\usepackage[colorlinks=true, citecolor=blue, urlcolor=blue, linkcolor=blue]{hyperref}
\newcommand{\vmax}{V_{\rm{max}}}

\newcommand{\msub}{M_{\rm{sub}}}

\newcommand{\mvir}{M_{\rm{vir}}}

\newcommand{\rvir}{r_{\rm{vir}}}
\newcommand{\vvir}{V_{\rm{vir}}}
\newcommand{\tvir}{T_{\rm{vir}}}

\newcommand{\mstar}{M_{\star}}
\newcommand{\msun}{M_{\odot}}

\newcommand{\lsun}{L_{\odot}}

\newcommand{\mpc}{{\rm Mpc}}

\newcommand{\kms}{{\rm km \, s}^{-1}}

\newcommand{\vpeak}{V_{\rm peak}}
\newcommand{\mpeak}{M_{\rm peak}}
\newcommand{\macc}{M_{\rm infall}}

\newcommand{\msunyr}{\msun\,{\rm yr^{-1}}}

\newcommand{\mcrit}{M_{\rm c}}
\newcommand{\mdotzero}{\dot{M}_{10}}
\newcommand{\mdot}{\dot{\mstar}}
\newcommand{\mab}{{\rm M_{UV}}}

\title[Near-field constraints on reionization]
{Near-field limits on the role of faint galaxies in cosmic reionization}
\author[Boylan-Kolchin, Bullock, and Garrison-Kimmel]
{Michael Boylan-Kolchin,$^1$\thanks{$\!\!$email: mbk@astro.umd.edu} James
  S. Bullock,$^2$ and 
  Shea Garrison-Kimmel$^2$\\
\noindent $\!\!^1$Department of Astronomy and Joint Space-Science Institute, 
  University of Maryland, College Park, MD 20742, USA\\
\noindent $\!\!^2$Center for Cosmology, Department of Physics and Astronomy, 
  4129 Reines Hall, University of California, Irvine, CA 92697, USA}
\begin{document}

 \pagerange{\pageref{firstpage}--\pageref{lastpage}} 
 \pubyear{2014}

\maketitle

\label{firstpage}
\begin{abstract}
  Reionizing the Universe with galaxies appears to require significant star
  formation in low-mass halos at early times, while local dwarf galaxy counts
  tell us that star formation has been minimal in small halos around us today.
  Using simple models and the {\tt ELVIS} simulation suite, we show that
  reionization scenarios requiring appreciable star formation in halos with
  $\mvir \approx 10^{8}\,\msun$ at $z=8$ are in serious tension with galaxy
  counts in the Local Group. This tension originates from the seemingly
  inescapable conclusion that 30 -- 60 halos with $\mvir > 10^{8}\,\msun$ at
  $z=8$ will survive to be distinct {\em bound} satellites of the Milky Way at
  $z = 0$.  Reionization models requiring star formation in such halos will
  produce dozens of bound galaxies in the Milky Way's virial volume today (and
  100 -- 200 throughout the Local Group), each with $\gtrsim 10^{5}\,\msun$ of
  {\em old} stars ($\gtrsim 13$ Gyr).  This exceeds the stellar mass function of
  classical Milky Way satellites today, even without allowing for the
  (significant) post-reionization star formation observed in these galaxies.
  One possible implication of these findings is that star formation became
  sharply inefficient in halos smaller than $\sim 10^9 \,\msun$ at early times,
  implying that the high-$z$ luminosity function must break at magnitudes
  brighter than is often assumed (at $\mab \simeq -14$).  Our results suggest
  that \textit{JWST} (and possibly even \textit{HST} with the Frontier Fields)
  may realistically detect the faintest galaxies that drive reionization.  It
  remains to be seen how these results can be reconciled with the most
  sophisticated simulations of early galaxy formation at present, which predict
  substantial star formation in $\mvir \sim 10^8 \, \msun$ halos during the
  epoch of reionization.
\end{abstract}


\section{Introduction} 
Star formation in the low redshift Universe is highly inefficient, both in a
galaxy-by-galaxy and in an ensemble-averaged sense. Less than 20\% of the cosmic
fraction of baryons in a typical Milky Way-mass dark matter halo has been
converted into stars by $z=0$ \citep{moster2010, guo2010, behroozi2013c}. The
conversion efficiency drops to much lower values in smaller halos: at
$\mvir=10^{10}\,\msun$, less than 1\% of a halo's baryonic allotment is in the
form of stars at the present day. This forms a stark contrast with the situation
needed for reionizing the Universe with star-forming galaxies, as current models
require high star formation rates in low-mass halos in order to initiate and
maintain reionization while also matching the WMAP/Planck optical depth to
electron scattering (e.g., \citealt{trenti2010, kuhlen2012a, robertson2013}).
Some estimates suggest that galaxies as faint as $\mab=-10$ may be required to
match the observed optical depth, which, given the high comoving number
densities implied, would demand that low-mass halos (with $\mvir \simeq 10^8
\,\msun$) are forming stars efficiently at these times \citep{schultz2014}.

\begin{figure*}
 \centering
 \includegraphics[scale=0.227, viewport=0 0 936 936]{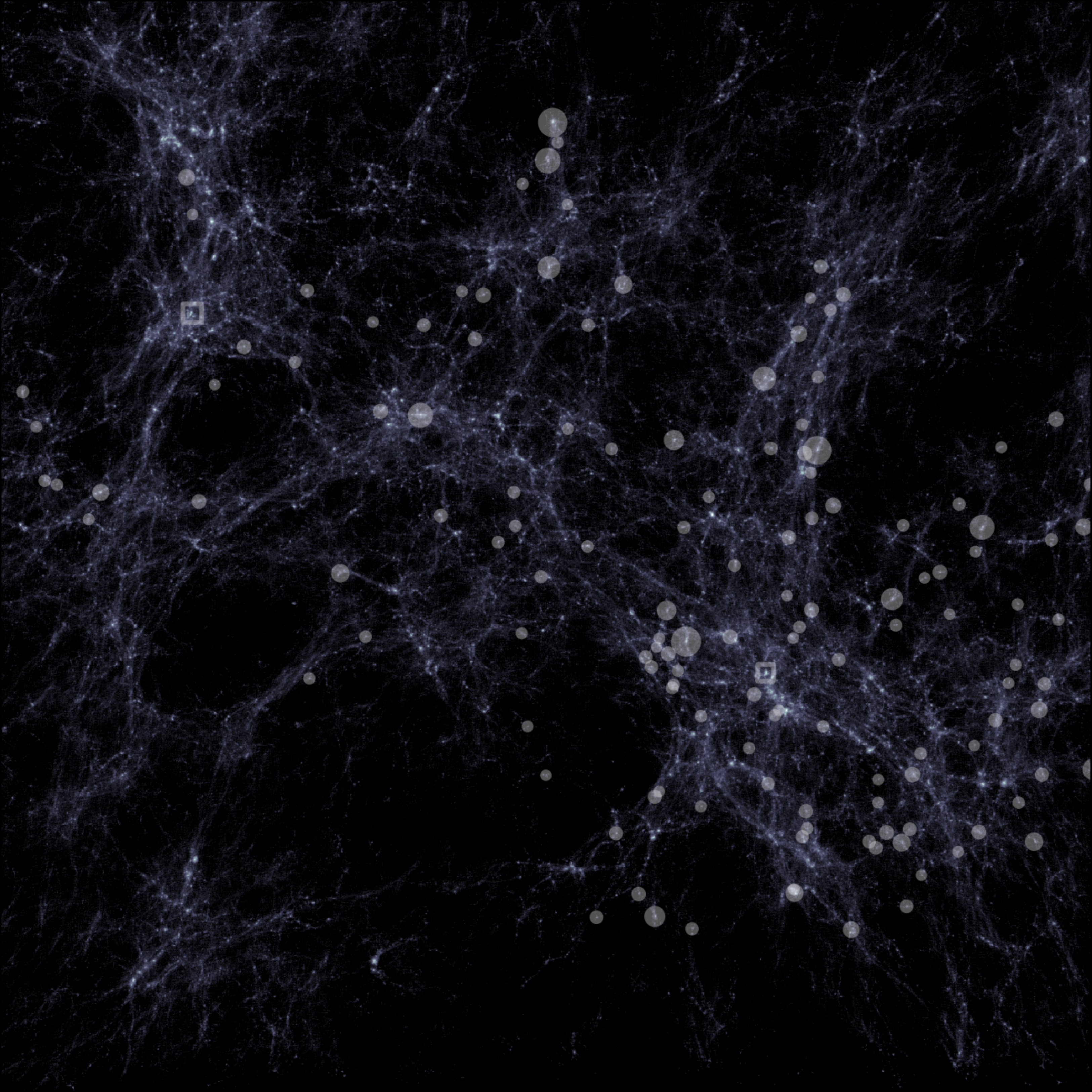}\,\,\,\,
 \includegraphics[scale=0.2, viewport=0 0 1098 1064]{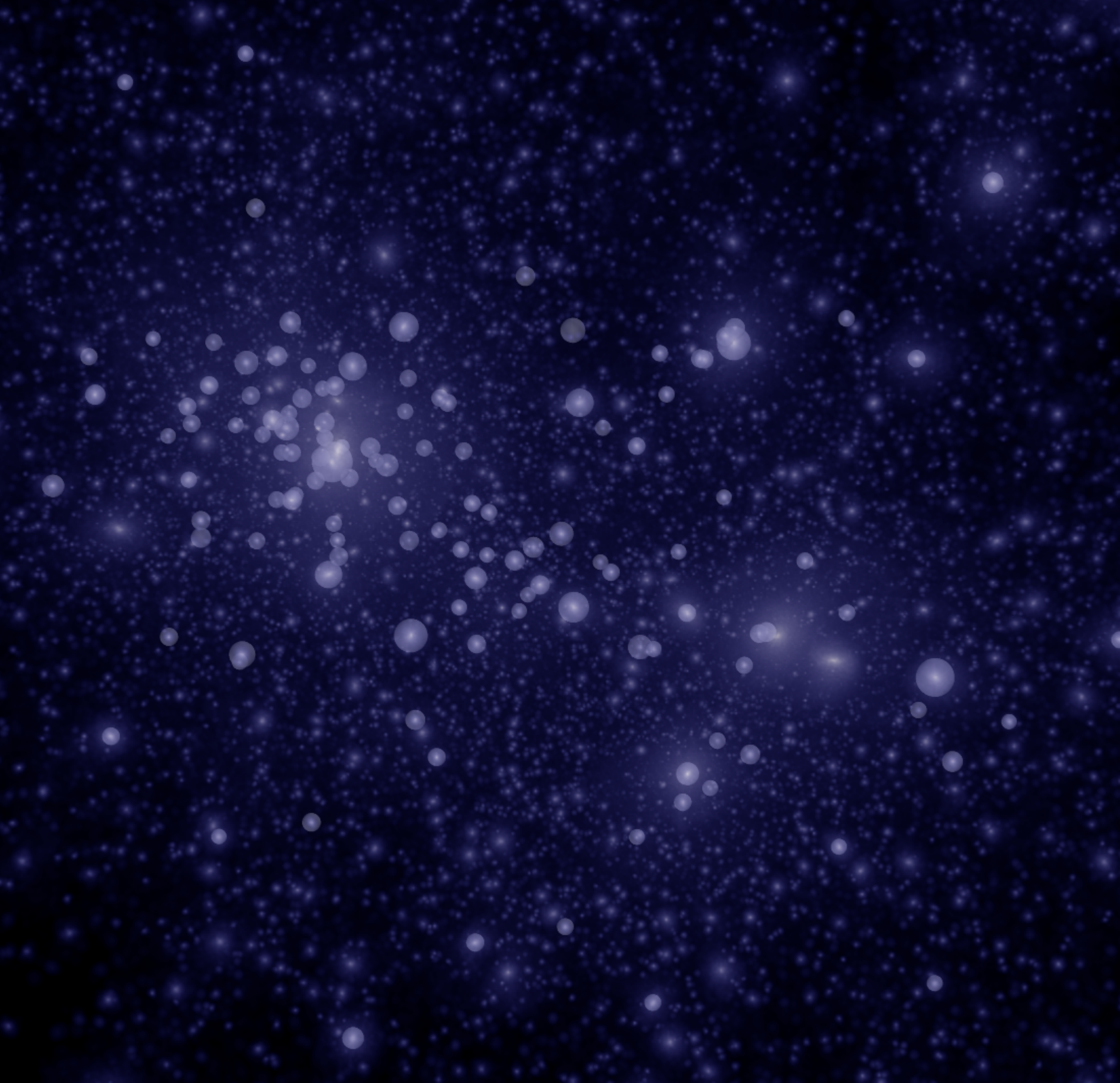}
 \caption{\textit{Left:} Density distribution at $z=8$ in a $10 \times 10\,{\rm
     comoving}\,\mpc$ region around the progenitor of a Local Group
   analog. Halos that have $\mvir(z=8)$ above the atomic cooling limit and that
   are the main branch progenitor of objects surviving to $z=0$ are marked with
   gray circles, while main branch progenitors of the MW/M31 analogs are marked
   with gray squares. \textit{Right:} Image of the same system at $z=0$ (1.5 x
   1.5 Mpc). Descendants of the $z=8$ halos marked in the left panel are shown
   with gray spheres, sized proportional to their $z=8$ virial radii.  There are
   approximately 140 identified objects in each image.
 \label{fig:image}
}
\end{figure*}

Numerical simulations predict that many halos that were massive enough to form
stars efficiently at high redshift have surviving, bound descendants in the
region surrounding the Milky Way (MW) today, indicating that there should be a
large number of potentially observable remnants of the reionization epoch in the
near field.  However, the small number of low-mass galaxies in our vicinity
strongly constrains the efficiency of star formation at high redshift, not only
in mini-halos \citep{madau2008}, but also in halos that are above the atomic
cooling limit of $\tvir \approx 10^4\,{\rm K}$, where cooling can proceed via
collisional line excitation of atomic hydrogen rather than relying on the
formation of ${\rm H}_2$. In this paper, we place constraints on the efficiency
and nature of star formation in low-mass halos during reionization by leveraging
the wealth of existing near-field data.

\section{Simulations, Data, and Models}
\label{section:methods}
To connect the Local Volume -- defined here as a sphere of radius 1.2 Mpc
centered on the Local Group barycenter -- to the high-redshift Universe, we use
the {\tt ELVIS} suite \citep{garrison-kimmel2014}, which is a set of zoom-in
simulations of Local Group (LG) analogs embedded in large high-resolution
regions. The primary simulations were performed in a WMAP7
cosmology\footnote{Specifically, the cosmological parameters are $\Omega_{\rm
    m}=0.266$, $\Omega_{\Lambda}=0.734$, $n_{\rm s}=0.963$, and $h \equiv
  H_0/(100\,\kms\,\mpc^{-1})=0.71$.}  \citep{komatsu2011} using particle masses
of $1.9 \times 10^{5}\,\msun$ in the high-resolution regions with a
Plummer-equivalent force softening of 141 pc. Halo catalogs in {\tt ELVIS} are
complete to $\msub=2\times10^7\,\msun$, or equivalently, $\vmax=8\,\kms$. The
convergence values in terms of ``peak'' quantities (defined along a halo's main
branch) are $\mpeak=6 \times 10^{7}\,\msun$ or $\vpeak = 12\,\kms$. This is
sufficient to resolve all halos above the atomic cooling limit of $\vvir \approx
17\,\kms$ or $\mvir (z=8) \approx 10^8\,\msun$. Merger trees are available for
all of the {\tt ELVIS} simulations; in this paper, we use a version of the trees
that tracks the most massive progenitor at each snapshot, which is slightly
different from what the standard trees provide. Further details about the {\tt
  ELVIS} suite can be found in \citet{garrison-kimmel2014}.

Fig.~\ref{fig:image} gives an initial visual impression of the matter
distribution around a LG analog, `Zeus and Hera', at $z=8$ (left) and $z=0$
(right). Massive ($\mvir > 10^8\,\msun$) objects at $z=8$ with an identifiable
descendant at $z=0$ are highlighted with gray spheres, sized proportional to
$\rvir(z=8)$ in both images. Only atomic-cooling halos at $z=8$ that are the
most massive progenitors of bound $z=0$ objects are marked, i.e., there is a
one-to-one correspondence between the circled halos in the left panel and those
in the right.  There are approximately 140 surviving, bound halos in and around
this LG analog that had main-progenitor masses above the atomic cooling limit at
$z=8$ and are therefore good potential sites for early galaxy formation.  In
subsequent sections, we quantify the number of massive $z=8$ halos surviving to
$z=0$ in the Local Volume and investigate the resulting implications for
reionization scenarios.

Maintaining cosmological reionization with galaxies requires a cosmological star
formation rate density (SFRD) that exceeds a critical value of
\begin{equation}
  \label{eq:1}
  \dot{\rho}_{\rm SFR} \approx 0.018 \msun \,{\rm yr}^{-1} \,\mpc^{-3} 
  \left( \frac{1+z}{8} \right)^3 \frac{C_3}{f_{\rm esc, 0.2}} \, T_4^{-0.845}
\end{equation} 
\citep{madau1999, shull2012}. This critical SFRD depends on the effective
clumping factor $C_{\rm H}\equiv \langle n_{p^+}^2 \rangle/\langle n_{p^+}
\rangle^2$ of ionized hydrogen ($C_3=C_{\rm H}/3$) and the escape fraction
$f_{\rm esc}$ of ionizing photons from galaxies ($f_{\rm esc, 0.2}=f_{\rm
  esc}/0.2$), as well as the temperature of the inter-galactic medium,
$T_{4}\equiv T_{\rm IGM}/10^{4}\,{\rm K}$. We assume that the star formation
rate is a function of halo mass, with
\begin{equation}
  \label{eq:2}
  \mdot = \mdotzero \,\left( \frac{\mvir}{10^{10} \, \msun}\right)^\beta \, 
\end{equation}
for halo masses above a critical value $\mcrit$ and $\mdot=0$ for $\mvir < \mcrit$.
We then have
\begin{equation}
  \label{eq:6}
   \dot{\rho}_{\rm SFR, sim}= \mdotzero \, \int_{\mcrit}^\infty\,\left(
     \frac{M}{10^{10} \, \msun}\right)^\beta \,\frac{dn}{dM}\,dM\,.
\end{equation}

\begin{figure}
 \centering
 \includegraphics[scale=0.5, viewport=10 0 495 420]{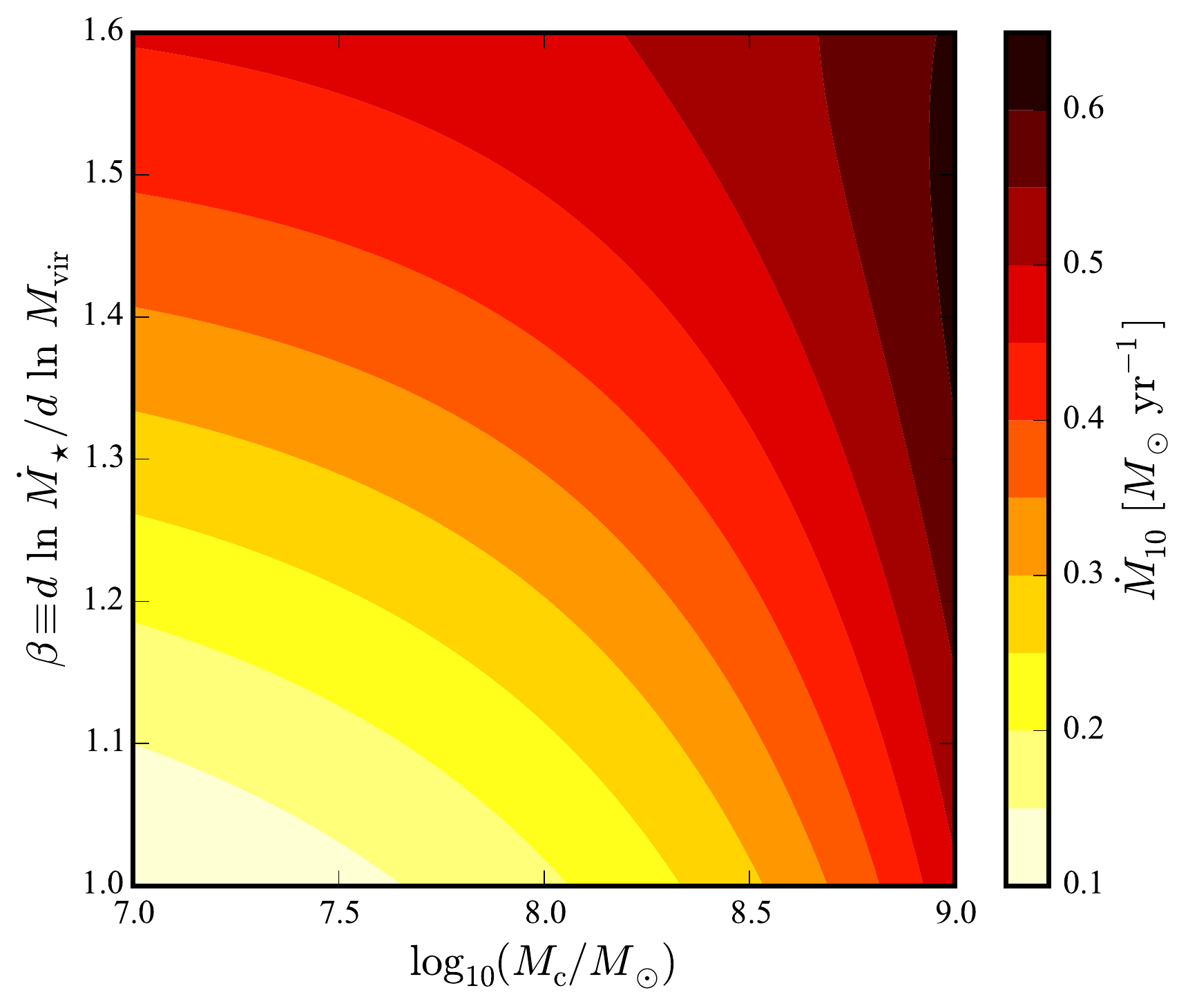}
 \caption{The star formation rate in $10^{10}\,\msun$ halos at $z=8$,
   $\mdotzero\,[\msun\,{\rm yr}^{-1}]$, required
   for reionization as a function of the minimum halo mass for star formation
   ($\mcrit$) and the slope of the $\mdot-\mvir$ relation, assuming a SFR-halo
   mass relation given by Eqn.~\ref{eq:2}. 
   For typical estimates of $\beta \sim 1.35$ and $\mdotzero \approx 0.4$, star
   formation in halos down to $\mcrit \approx 10^{8}\,\msun$ is required.
   Steeper $\mdot-\mvir$ relations (larger values of
   $\beta$) require larger values of $\mdotzero$ at fixed $\mcrit$ to
   achieve reionization.
 \label{fig:sfr_density}
}
\end{figure}

For any value of $\mcrit$ and $\beta$, we can determine the required
normalization $\mdotzero$ of the star formation rate to reionize the Universe
with galaxies through a comparison with Eqn.~\ref{eq:1}.
Figure~\ref{fig:sfr_density} shows contours of the required $\mdotzero$ (in
units of $\msunyr$) at $z=8$ in $\mcrit-\beta$ space, adopting the mass function
for dark matter halos given by \citet*{sheth1999} [which, we have confirmed,
gives a good match to the halo abundance in the simulations of
\citet{schultz2014} and \citet{vogelsberger2014}]. \citet{finlator2011} find
$\beta\approx 1.35$ and $\mdotzero=0.42\,\msunyr$ at $z=8$ in their simulations,
giving $\mcrit \approx 1.5 \times 10^{8}\,\msun$ (which is just at their
resolution limit) -- i.e., star formation is required in all halos above the
atomic cooling limit in order to achieve reionization in these simulations.  For
steeper values of $\beta$ in the $\mdot-\mvir$ relation, higher normalizations
$\mdotzero$ and/or lower cut-off masses $\mcrit$ are required.

\section{Connection to Redshift Zero}
\label{sec:z0connect}

\begin{figure}
 \centering
 \includegraphics[scale=0.5, viewport=10 0 475 420]{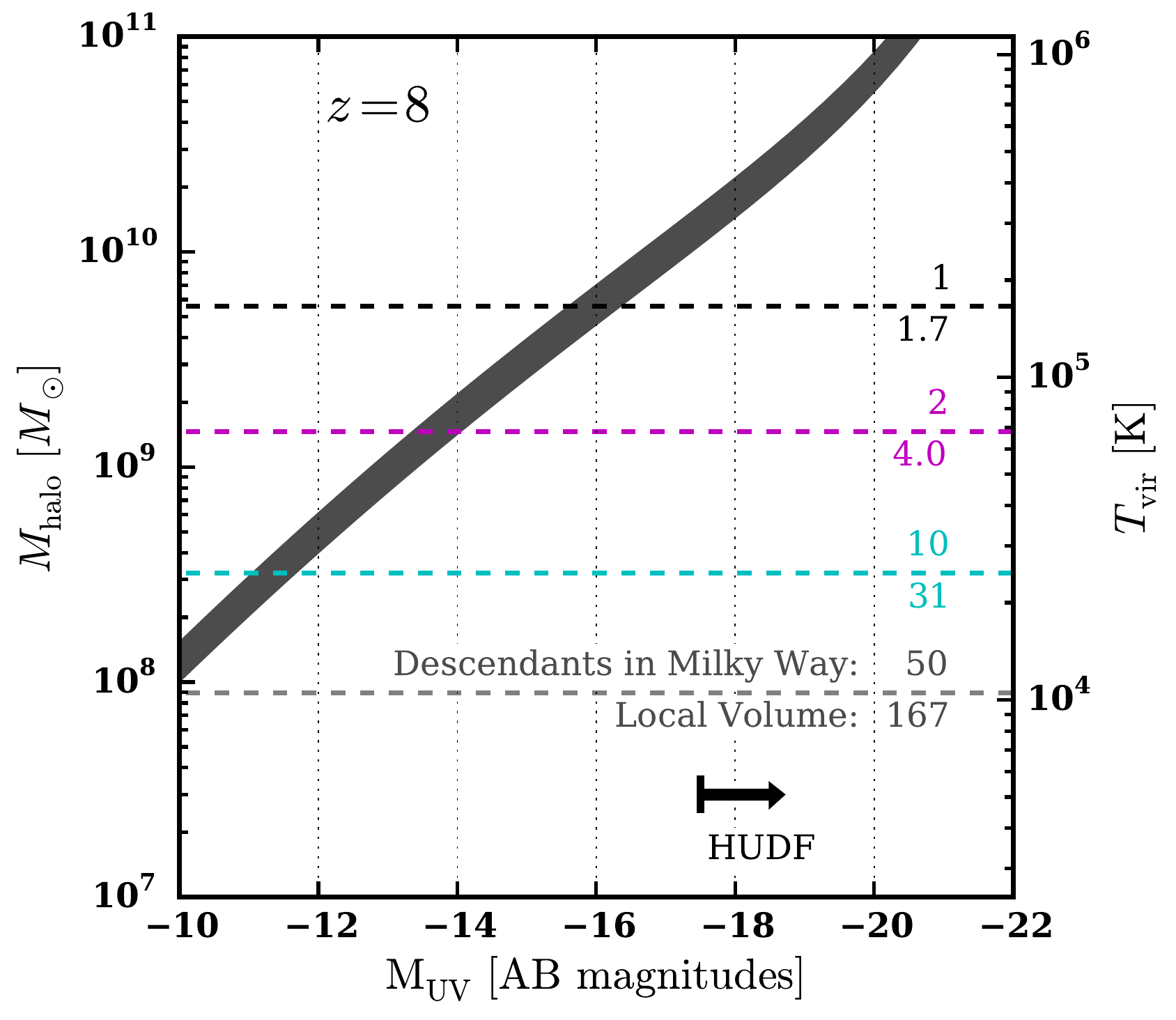}
 \caption{The abundance matching relation between UV luminosity and halo mass at
   $z=8$ (thick black curve), which is well-approximated by Eqn.~\ref{eq:2} with
   $\mdot=0.33\,\msunyr$ and $\beta=1.45$. Also given are counts of
   identifiable descendant halos at $z=0$ above a given $z=8$ mass inside both
   the simulated MW and Local Volume. There should be approximately 50 surviving
   remnants in the present-day Milky Way of halos that had masses above the
   atomic cooling limit, which are predicted to host $\mab (z=8) \approx -10$
   galaxies. The main progenitor of the Milky Way itself has $\mab \approx -16$
   and is hosted by a halo of $\tvir \approx 1.5\times10^{5}\,{\rm K}$ ($\mvir
   \approx 6 \times 10^{9}\,\msun$).
 \label{fig:z8sham_counts}
}
\end{figure}

In order to schematically associate dark matter halo masses with observed UV
luminosity functions at $z=8$, we use abundance matching \citep{conroy2006}. As
in Sec.~\ref{section:methods}, we use the Sheth-Tormen halo mass function; we
take the UV luminosity function measured by \citet{schenker2013}, which is very
similar to that of \citet{schmidt2014} and \citet{bouwens2014}. As the Hubble
UDF is only complete for $\mab \la -17.5$ at $z=8$, we extrapolate the
\citet{schechter1976} fit of Schenker et al. to lower luminosities.

The result, shown in Fig.~\ref{fig:z8sham_counts}, can be directly related to
our parameter choices in Eqn. 2 and Fig. 2.  Assuming $L_{\rm UV} \propto \mdot
\propto \mvir^\beta$, our abundance matching relation gives $\beta \simeq 1.4$.
The normalization of the mapping from $L_{\rm UV}$ to $\mdot$ depends on
metallicity and the assumed stellar initial mass function (IMF). For the
fiducial value of $\mathcal{K}_{\rm FUV}=1.15\times10^{-28}\,\msunyr$ given in
\citet{madau2014}, we obtain
\begin{equation}
  \label{eq:7}
\log_{10}\left(\frac{\mdot}{\msunyr}\right)=-0.4\,(\mab+18.22)\,,
\end{equation}
which, combined with the relation shown in Fig.~\ref{fig:z8sham_counts}, gives
$\mdotzero=0.33\,\msunyr$ for the normalization of Eqn.~\ref{eq:2}. Allowing for
variations with age/metallicity gives a range of $0.28 \la \mdotzero/(\msunyr)
\la 0.46$, while changing the assumed IMF from Salpeter to Chabrier or Kroupa
reduces this normalization by a factor of $1.5-1.6$ \citep{madau2014};
additional effects such as stellar rotation may also reduce the star formation
rate at fixed UV luminosity by 30\% \citep{horiuchi2013}. Combined, these
effects lead to a factor of $\sim 2-3$ uncertainty in the normalization of
Eqn.~\ref{eq:2}, which is not enough to have a qualitative effect on the results
discussed in the following sections.

According to Fig.~\ref{fig:z8sham_counts}, $\mab=-10$ corresponds to the atomic
cooling limit of $\mvir \approx 10^{8}\,\msun$, while halos with $\mvir \approx
10^{9}\,\msun$ typically host galaxies with $\mab=-13$. The faintest $z=8$
galaxies in the HUDF have $\mab \approx -17.5$, corresponding to a halo mass of
$\approx 10^{10}\,\msun$. Figure~\ref{fig:z8sham_counts} also indicates, for a
variety of $z=8$ halo masses, the average counts of surviving descendants at
$z=0$ within 300 kpc of a Local Group giant (labeled `Milky Way') or within 1.2
Mpc of the LG barycenter (labeled `Local Volume'). The main progenitor of the
Milky Way is predicted to have a virial mass of approximately $6\times
10^9\,\msun$ ($\tvir \approx 1.5 \times 10^5\,{\rm K}$) at $z=8$, corresponding
to $\mab=-16$.

\begin{figure}
 \centering
 \includegraphics[scale=0.56, viewport=15 0 400 460]{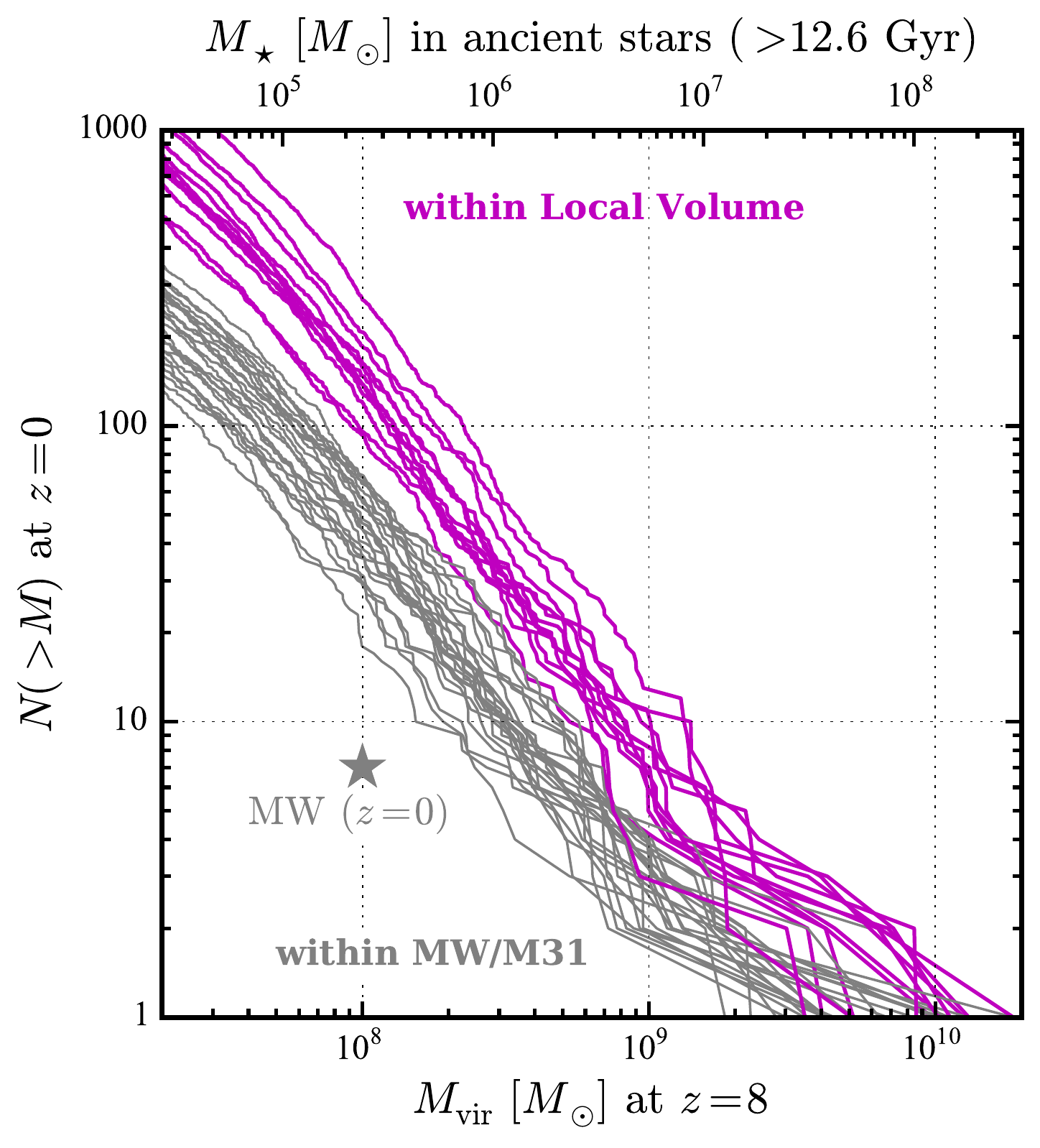}
 \caption{Mass functions at $z=8$ of the most massive progenitors of dark matter
   halos within 300 kpc of the MW (gray) or within the Local Volume (magenta) at
   $z=0$. The upper horizontal axis gives the mass in old stars ($>\!12.6$ Gyr)
   associated with each of these bound halos at $z=0$ (see text for details).
   This reionization model predicts 20--50 bound galaxies in the MW's virial
   volume today that have $> 3 \times 10^{5}\,\msun$ of stars that formed before
   $z = 6$.  There are only twelve galaxies with this stellar mass in the Milky
   Way \textit{at the present day}, almost all of which formed most of their
   stars after $z=6$. Only 6--8 MW satellites (gray star) have $\mstar \gtrsim
   2\times10^{5}\,\msun$ in ancient stars \citep{weisz2014}.
   \label{fig:z8_smf}
 }
\end{figure}

The connection between atomic cooling halos in the epoch of reionization and
their local remnants is presented in more detail in Figure~\ref{fig:z8_smf}. The
gray curves show cumulative mass functions for the most massive progenitors at
$z=8$ of objects surviving to $z=0$ within a MW/M31 analog, while the magenta
curves show the same quantity within the Local Volume. The upper horizontal axis
of the Figure shows the stellar mass in old stars ($>12.6\,{\rm Gyr}$)
associated with each surviving galaxy assuming a star formation rate at $z=8$
given by Eqn.~\ref{eq:2} with $\mdotzero=0.42$ and $\beta=1.36$, as found in
\citet{finlator2011}, and a star formation duration of $\Delta_{\rm t,
  sf}=300\,{\rm Myr}$ [the time between $z=8$ and $z=6$, or equivalently,
$0.3\,t_{\rm Hubble}(z=8)$]: $\mstar = \Delta_{\rm t, sf}\,\mdotzero\,
(\mvir/10^{10}\,\msun)^{\beta}$. These stellar masses can be easily scaled for
different assumptions about $\Delta_{\rm t, sf}$. This calculation assumes that
there was no star formation prior to $z=8$; virtually all models predict star
formation starting earlier than this in halos at (and below) the atomic cooling
limit, which would increase our estimates of the mass contained in ancient
stars.

Figure~\ref{fig:z8_smf} emphasizes the tension between requirements for
reionization and the low number of observed satellite galaxies around the MW
today. \textit{Within the virial volume of the Milky Way at $z=0$, there should
  be $\sim 50$ distinct remnants of halos that had, at redshift eight, virial
  masses exceeding the atomic cooling limit; the number grows to $\sim 170$ when
  considering the Local Volume.} For models that require star formation in halos
at the atomic cooling limit to produce enough ionizing photons through star
formation to maintain reionization at these redshifts, a star formation duration
of only 100 Myr (rather than the 300 Myr assumed in the Figure) already
overproduces the stellar mass function of Local Group galaxies today --- 13
galaxies with $\mstar > 2 \times 10^{5}\,\msun$ in the Milky Way, twice that in
M31, and $\sim 50$ in the Local Volume (e.g., \citealt{mcconnachie2012}) ---
even though much of the star formation in observed MW satellites happens at
significantly later times\footnote{We have not considered stellar mass loss,
  which could, in principle, act to somewhat reduce the stellar
  mass between $z=7$ and $z=0$. However, this effect should not be very large,
  and we expect will be more than compensated for by additional star formation
  before $z=8$ and after $z=6-7$ and by mergers.}  \citep{weisz2014}.

\section{Discussion}
The competing requirements of efficient star formation in low-mass halos at high
redshift to initiate and maintain reionization versus inefficient star formation
in low-mass halos (integrated over cosmic time) to match the low number of such
halos in and around the Local Group at present strongly constrain reionization
scenarios, insofar as the LG is in any way representative. Models that extend to
$\mcrit \approx 10^{8}\,\msun$ -- approximately $\mab=-10$, according to our
abundance matching relation -- will over-produce the number of galaxies having
$\mstar > 10^5\,\msun$ at present (and will do so with only $\sim 100$ million
years of star formation at $z=8$).  To avoid this tension, models must increase
the critical mass below which star formation is heavily suppressed to $\mcrit
\sim 10^9 \msun$; to still produce enough star formation to reionize the
Universe, this increase must be accompanied by an increase in $\mdotzero$, an
increased ionizing photon output per galaxy, or a shallower SFR-halo mass slope
$\beta$.  Note that relations with $\beta < 1.4$ would necessarily imply that
the UV luminosity function must steepen beyond its already steep value of
$\alpha \approx -1.9$ \citep{schenker2013, bouwens2014, schmidt2014} at
luminosities below the HUDF limit.

\subsection{Comparison with previous results}
At present, the most sophisticated numerical simulations of reionization-era
galaxy formation find high stellar masses ($\ga 10^{5}\,\msun$) in halos at or
below the atomic cooling limit in the high-redshift universe, supporting the
tension discussed this paper. \citet{wise2012a} show that radiation pressure is
important in regulating the star formation in a high-resolution simulation of a
$z=8$ dwarf; their simulation produces $\mstar=4.5\times10^{5}\,\msun$ in a halo
mass of $2.0\times 10^{8}$ at $z=8$. In a follow-up paper, \citet{wise2014} find
that the typical stellar mass of a $10^{8.5}\,\msun$ dark matter halo at
$z\sim8$ is $10^{5.74}\,\msun$, which is very similar to our simple estimate in
Figure~\ref{fig:z8_smf}.  \citet{simpson2013} simulate a halo with $\mvir(z=0)
\approx 10^9\,\msun$ at very high mass and spatial resolution and find that it
has formed $10^5\,\msun$ of stars by $z=8$, when the halo has a virial mass of
$\sim 2 \times 10^7\,\msun$.

Figure~\ref{fig:z8_smf} indicates that the Milky Way should contain tens to
hundreds of such dwarf satellites today for the Wise et al. and Simpson et
al. results, respectively, meaning there should be this number of dwarf galaxies
around the MW even without any star formation subsequent to $z \sim 8$. In
reality, there are only $\sim 12$ (30) known satellites of the MW (M31) with
$\mstar > 10^{5}\,\msun$ at $z=0$, and all of these have significant star
formation after $z=6$ \citep{weisz2014}. It appears difficult to avoid the
conclusion that typical halos of $\sim 10^{8}\,\msun$ at $z=8$ must be forming
stars inefficiently, with star formation rates that are considerably lower than
the $\mdot \approx 0.001-0.01 \,\msunyr$ found in many current simulations and
models -- meaning they should not be major contributors to reionization.

Our results therefore argue for a critical mass closer to $10^{9}\,\msun$ rather
than $10^8\,\msun$ in order to avoid over-producing the observed abundance of
classical ($L_V>10^{5}\,\lsun$) satellites of the Milky Way (and doing so by $z
\sim 7$, without any subsequent star formation). A halo mass of $10^{9}\,\msun$
corresponds to $\mab \approx -13$ via Figure~\ref{fig:z8sham_counts}, a value
that is at the edge of the allowed values for simultaneous achieving
reionization with galaxies and matching the optical depth to electron
scattering, according to \citet{robertson2013}. 

Our results are complementary to those of \citet{madau2008}, who focused on the
star formation in mini-halos $10^{6} \la \mvir/\msun \la 10^{8}$ and showed that
star formation efficiencies are limited to $\la 10^{-3}$ to avoid overproducing
the abundance of ultra-faint dwarfs. Our analysis focuses on somewhat larger
mass scales, potentially comparable to progenitors of the classical MW dwarfs,
and we find a qualitatively similar result. This concern was also raised by
\citet*{bovill2011a}: they used a hybrid of hydrodynamical and collisionless
simulations and also found their models over-produced the number of predicted MW
satellites having $L_V > 10^{4}\,\lsun$. A novel direction of our work is to
combine limits derived from abundances of descendants of $z=8$ halos in the MW
with constraints coming from models of reionization. Specifically, we point out
that the dwarfs at $z=0$ might be the same as the ones required for reionization
and quantify the resulting implications. The large high-resolution region of
each of the {\tt ELVIS} simulations also enables us to study the distribution of
descendants throughout the Local Volume, not just within the virial radius of
the Milky Way. In the future, it will be interesting to compare the
simulation-based models we have presented here with the detailed extended
Press-Schechter (\citeyear{press1974}) models of, e.g., \citet*{munoz2011} and
\citet{salvadori2014}. The interconnection between high-redshift star formation
in atomic cooling halos, globular cluster formation, and reionization
\citep{boley2009}, along with the spatial location of the remnants of such halos
around the Milky Way today \citep{gao2010}, is likely to be a fruitful direction
for further study.

\subsection{Further Implications}
Our results provide mass scales at $z=8$ for expected hosts of MW satellites and
for the main progenitor of the Milky Way itself. Approximately fifty distinct
progenitors with $z=8$ masses above $10^8\,\msun$ survive to end up within 300
kpc of the MW at $z=0$. \textit{A robust result of our analysis is that the main
  progenitors of classical MW satellites had masses above the atomic cooling
  limit at $z=8$.}  If atomic cooling halos play a major role in reionization,
the classical MW dwarfs are descendants of these crucial galaxies.
  
We find that the main progenitor of the Milky Way itself was likely to have a
mass of $6\times 10^{9}\,\msun$ at $z=8$; from the scatter in our simulations,
this number could be a factor of three higher or lower. At the high end of this
range, Fig.~\ref{fig:z8sham_counts} indicates that it would be at the edge of
detectability in the HUDF, and well in the range accessible by the Frontier
Fields campaign \citep{coe2014}.

We note that all of our results only take into account the most massive
progenitor of $z=0$ halos. In reality, many halos at $z=0$ have multiple
progenitors at $z=8$, meaning we have only provided a lower limit on the stellar
mass functions in ancient stars we expect to see in the LG today. If we consider
\textit{all} progenitors of everything within the simulated Local Volume at
$z=0$, we find that approximately 50\% of progenitor halos with $\mvir(z=8) >
10^{8}\,\msun$ end up merging into the main halo of the Milky Way or M31 while
50\% are progenitors of other Local Volume objects at $z=0$. A further relevant
scale of note is the typical peak mass $\macc$ of halos with a given mass at
$z=8$; we find $\macc/\mvir(z=8) \approx 5-20$ with a median value of $\sim 10$
(see also fig. 4 of \citealt{boylan-kolchin2012}), indicating that the virial
masses of atomic cooling halos grow considerably after $z=8$. This is also
consistent with the aforementioned results of Simpson et al.

Wyithe and Loeb (\citeyear{wyithe2013, wyithe2014}) have argued for a relatively
low duty cycle ($\la 0.1$) of star formation in high-redshift galaxies.  The
abundance matching results of Figure~\ref{fig:z8sham_counts} would need to be
modified in this scenario. In particular, since only a fraction of halos would
host UV-luminous galaxies in this model, the abundance matching curve would
shift to the right such that galaxies of a fixed UV luminosity would be hosted
by less massive dark matter halos. If the duty cycle is 10\% irrespective of
halo mass, then halos with $\mvir(z=8)=10^8\,\msun$ would be expected to host
galaxies with $\mab \approx -13$. Although 90\% of halos at a given mass would
be non-star-forming at any time in this model, the ones that are forming stars
do so at a much higher rate than we have assumed above. The product of
$\Delta_{\rm t, sf}$ and $\mdotzero$ is therefore comparable or even larger than
what we have assumed in Sec.~\ref{sec:z0connect}, once again substantially
over-producing the satellite luminosity function of the Milky Way at the present
day (see also fig.~6 of \citealt*{wyithe2013}).

Star formation in halos of $\mvir=10^{8}-10^{9}\,\msun$ is expected to be
sensitive to the presence of an ionizing background. If the Local Volume was
reionized well before $z=8$, our conclusions would be altered. One
interpretation of our results, then, is that Milky Way was indeed reionized
early, leading to more negative feedback in low-mass halos destined to end up in
the Milky Way and effectively raising $\mcrit$ in the Milky Way without doing so
in the Universe on average. If this is not the case, our results imply that
forthcoming deep-field observations with \textit{HST} in the Frontier Fields and
with \textit{JWST} may access the smallest galaxies that are important
contributors to reionization.

\vspace{-0.2cm}
\section*{Acknowledgments} 
We thank Manoj Kaplinghat and Massimo Ricotti for helpful conversations. Support
for this work was provided by NASA through a \textit{Hubble Space Telescope}
theory grant (program AR-12836) from the Space Telescope Science Institute
(STScI), which is operated by the Association of Universities for Research in
Astronomy (AURA), Inc., under NASA contract NAS5-26555. This work was also
supported by a matching equipment grant from UC-HiPACC, a multicampus research
program funded by the University of California Office of Research.

\bibliography{draft_final.bbl}

\label{lastpage}
\end{document}